# Extended Impedance Modal Analysis of Internal Dynamics in Grid-Following Inverters

Le Zheng, *Member*, *IEEE*, Jiajie Zheng, Lei Chen, *Senior Member*, *IEEE*, and Chongru Liu, *Senior Member, IEEE*

*Abstract*—**As the penetration of Grid-Following inverters (GFL) in power systems continues to increase, the dynamic characteristics of power systems undergo significant transformations. Recently, modal analysis based on the impedance model (MAI) has been utilized to evaluate the interaction between GFLs and the power grid through impedance/admittance participation factors (PF). However, MAI relies on the impedance model that characterizes the input-output behavior of the ports, treating the GFL as a single integrated entity. This approach limits its ability to reveal the complex dynamic coupling mechanism between different control loops inside the GFL. In this paper, we propose an extended impedance modal analysis (EMAI) method. Firstly, the equivalent dynamics of the GFL are decomposed into two components: the synchronous dynamic dominated by the phase-locked loop (PLL) and the electromagnetic dynamic dominated by the current control loop (CCL). The PF and participation ratios (PR) of these two dynamic components are then calculated to identify the dominant dynamics of the system. Building on this, we introduce the explicit parameter participation factors (PPF) to further pinpoint the key parameters of the dominant control loop, which provides a way for enhancing system stability. Finally, the effectiveness of the proposed method is validated through the simulation of the modified 14-bus and 68-bus systems. The EMAI method enables the analysis of the dynamic characteristics of each control loop in the GFL based on the impedance model. It can effectively identify the critical control loops influencing system stability without requiring the construction of a full state-space model, demonstrating its broad applicability and value.**

*Index* **Grid-Following inverter, admittance decomposition, participation factor, extended impedance modal analysis.**

## I. INTRODUCTION

The large-scale integration of Grid-Following inverters (GFL) has significantly altered the dynamic characteristics of power systems, raising widespread concerns about instability resulting from inverter-grid interactions [1]-[2]. GFLs are reshaping system dynamics while reducing grid strength and inertia [3]-[4]. At the same time, the interactive coupling among controllers operating on different time scales within GFLs complicates stability analysis [5]. Therefore, it is urgent to develop new analytical frameworks to facilitate detailed analysis of the dynamic characteristics of each control loop in GFLs, addressing the increasingly complex stability challenges [6]-[7].

Modal Analysis Based on the State-space Model (MASS) is an important method to identify the critical factors affecting the system stability. MASS employs the Participation Factor (PF) to quantify the contribution of each state variable to a specific mode [8]. However, the proliferation of electrical elements in new power systems has greatly increased the complexity of state-space modeling for the entire system [9]. In addition, state-space modeling requires detailed knowledge of system topology and complete control parameters for each element. In practice, inverters are often characterized by impedance models that describe the voltage and current ports behaviors, typically with gray-box or black-box characteristics [10]. Therefore, the application of MASS for stability analysis in new power systems with GFLs needs further investigation.

Several studies have investigated methods for utilizing impedance models to identify the critical factors contributing to oscillations. Ref. [11] introduces the Resonance Mode Analysis technique to pinpoint the bus with the highest participation. Eigenvalue sensitivity has been derived and applied to determine the pivotal network elements influencing some specific oscillatory modes [12]-[13]. Loop/node participation factors, defined by the frequency domain matrix, are used to analyze the effect of oscillations [14]. Additionally, eigenvalue trajectories have been employed to evaluate the impact of controller parameters on stability [15]. Ref. [16] indicates the necessity of considering not only the amplitude-frequency response of characteristic values but also their quality factor when analyzing resonance modes.

Recently, system stability has been assessed by embedding frame dynamics into the whole-system dynamic matrix [17]. Building on this, Modal Analysis based on the Impedance model (MAI) can evaluate the contribution of individual devices to oscillation modes at the device level [18]-[19]. In addition, the corresponding parameter participation factors (PPF) provide insight for improving system damping. However, unlike MASS, MAI treats the inverter as a single integrated element, which limits its ability to pinpoint the primary causes of instability at the control loop and state variable levels.

It is of great significance to study the root cause of system instability in GFLs to improve system stability. A method for bidirectional mapping of electrical and mechanical ports

Manuscript received xx; revised xx. This work was supported by the National Natural Science Foundation of China under Grant 52307095. (*Corresponding author: Chongru Liu*, e-mail: liu.chongru@ncepu.edu.cn).

Le Zheng, Jiajie Zheng, and Chongru Liu are with the State Key Lab of Alternate Electrical Power System with Renewable Energy Sources, North China Electric Power University, Beijing 102206, China. Lei Chen is from Department of Electrical Engineering, Tsinghua University, Beijing 100084, China.

provides a new perspective for analyzing stability arising from different factors [20]. In addition, separating the dynamics within GFLs by introducing distinct input and output variables for different ports offers another valuable approach. However, this method is limited by the practical challenges of obtaining impedance information for various ports [21]. The decomposition of control loops into equivalent circuit elements enables the analysis of internal dynamics within the inverters [22]. Nevertheless, further exploration is needed to determine how to effectively decompose the equivalent admittance of different dynamics within GFLs in the impedance model.

The Extended Impedance Modal Analysis (EMAI) is introduced in this paper to analyze the influence of different dynamics within the GFLs and to identify the root causes of instability in complex power systems. The main contributions of this paper are as follows:

(1) An admittance decomposition algorithm based on the matrix inversion lemma is proposed, separating the dynamics of a GFL into synchronous dynamics (SD), governed by the phase-locked loop (PLL), and electromagnetic dynamics (ED), primarily governed by the current control loop (CCL). This dynamics decomposition enables an in-depth exploration of the interactive coupling effects among various GFL control loops at different time scales.

(2) Overall PF and Participation ratios (PR) for different dynamics are proposed to evaluate the contributions of SD and ED at the control loop level, enabling root cause tracking of system instability. Furthermore, the explicit PPF is introduced to pinpoint the key parameters of the dominant control loop, which can be utilized as an index to optimize the control parameter and enhance system damping.

The structure of this paper is as follows: Section II summarizes the advantages and disadvantages of the MASS and MAI methods. Section III details the proposed admittance decomposition method for GFLs. In section IV, admittance PF, PR, and explicit PPF are proposed to characterize the participation of different dynamics. The effectiveness of the proposed EMAI method is demonstrated through case studies on modified 14-bus and 68-bus systems in Section V. Finally, Section VI concludes the work.

## II. MODAL ANALYSIS METHODS

### A. Modal Analysis Based on the State-Space Model

The linearized state-space equation of a power system can be expressed as follows:

$$\begin{cases} \Delta \dot{x} = A\Delta x + B\Delta u \\ \Delta y = C\Delta x + D\Delta u \end{cases} \tag{1}$$

where $\Delta x$ is the state vector, $\Delta u$ is the input vector, and $\Delta y$ is the output vector. $A$, $B$, $C$, and $D$ are the state matrix, input matrix, output matrix, and feedforward matrix, respectively.

The diagonalization of the state matrix $A$ by the $\Delta x = \Phi \Delta z$ coordinate transformation yields a diagonal state matrix.

$$\Lambda = \Psi A \Phi = \text{diag}(\lambda_1 \cdots \lambda_i \cdots \lambda_t) \tag{2}$$

where $\lambda_i$ is the $i$-th eigenvalue of matrix $A$, $\phi_i$ and $\psi_i$ in $\Phi = [\phi_1 \cdots \phi_i \cdots \phi_t]$ and $\Psi = [\psi_1^T \cdots \psi_i^T \cdots \psi_t^T]^T$ are the right eigenvector and left eigenvector of $\lambda_i$, respectively. We have $\Phi \Psi = E$, where $E$ is the unit matrix. $t$ is the dimension of the system state vector.

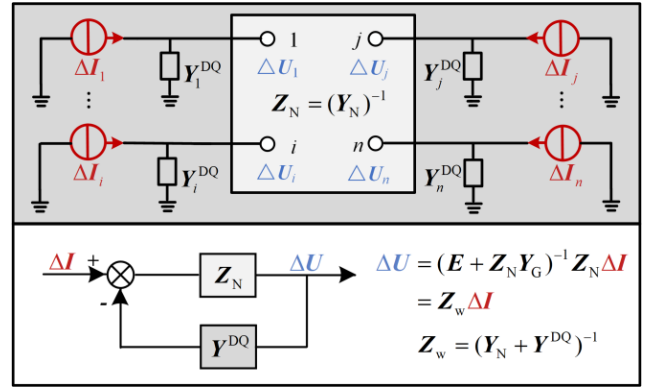

Fig. 1. Illustration of the whole-system impedance model.

The sensitivity of $\lambda_i$ to the $k$-th row and $j$-th column element $a_{kj}$ of the matrix $A$ is [8]:

$$\partial \lambda_i / \partial a_{kj} = \psi_{ik} \phi_{ji} \tag{3}$$

where $\phi_{ji}$ and $\psi_{ik}$ are the $j$-th element of the right eigenvector $\phi_i$ and the $k$-th element of the left eigenvector $\psi_i$, respectively.

The participation matrix $P = [P_1 \cdots P_i \cdots P_t]$ is as follows.

$$P_i = \begin{bmatrix} p_{1i} \\ \vdots \\ p_{ki} \\ \vdots \\ p_{ti} \end{bmatrix} = \begin{bmatrix} \phi_{1i} \psi_{i1} \\ \vdots \\ \phi_{ki} \psi_{ik} \\ \vdots \\ \phi_{ti} \psi_{it} \end{bmatrix} = \begin{bmatrix} \partial \lambda_i / \partial a_{11} \\ \vdots \\ \partial \lambda_i / \partial a_{kk} \\ \vdots \\ \partial \lambda_i / \partial a_{tt} \end{bmatrix} = \text{diag}(\frac{\partial \lambda_i}{\partial A}) \tag{4}$$

where $p_{ki}$ denotes the relative participation of the $k$-th state variable in the $i$-th mode, which equals the sensitivity of $\lambda_i$ to the $k$-th diagonal element $a_{kk}$ of matrix $A$.

MASS relies on detailed information of the entire system to compute the eigenvalues and participation factors, which is often limited by the black-box or gray-box nature of inverter and network models. Additionally, the curse of dimensionality and poor scalability of MASS in large-scale power systems further reduce its practicality.

### B. Modal Analysis Based on the Impedance Model

Considering the small-signal impedance model of the entire system in Fig. 1, where the total number of buses is $n$, and $Z_N$ represents the network nodal impedance matrix. Before connecting the impedance models, their respective coordinate systems must be aligned to the global coordinate system [17]. In the global DQ coordinate system, the admittance and impedance of the power source connected to bus $m$ are denoted as $Y_m^{DQ}$ and $Z_m^{DQ}$, respectively.

$$Y_m^{DQ} = \begin{bmatrix} Y_{mDD} & Y_{mDQ} \\ Y_{mQD} & Y_{mQQ} \end{bmatrix} = (Z_m^{DQ})^{-1} \tag{5}$$

In (5), $Y_{mDD}$, $Y_{mDQ}$, $Y_{mQD}$, and $Y_{mQQ}$ represent the four components of the admittance in the system's global DQ coordinate system. $Y^{DQ} = \text{diag}(Y_1^{DQ} \cdots Y_m^{DQ} \cdots Y_n^{DQ})$ is the admittance matrix of all power sources.

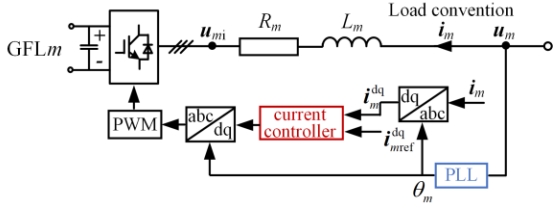

Fig. 2. GFL control structure.

Based on the closed-loop feedback relationship shown in Fig. 1, the whole-system dynamic impedance matrix $\mathbf{Z}_w$ can be expressed as:

$$\mathbf{Z}_w = (\mathbf{E} + \mathbf{Z}_N \mathbf{Y}^{DQ})^{-1} \mathbf{Z}_N \qquad (6)$$

The overall PF of power source $m$ in the $i$-th eigenvalue $\lambda_i$ of the system is defined as $\mathrm{PF}_m$ [18]-[19]:

$$\begin{cases} \Delta\lambda_i = \left\langle -\operatorname{Res}_{\lambda_i} \mathbf{Z}_{wm}^*, \Delta\mathbf{Y}_m^{DQ}(\lambda_i) \right\rangle \\ \quad = \varepsilon_m \left\langle -\operatorname{Res}_{\lambda_i} \mathbf{Z}_{wm}^*, \mathbf{Y}_m^{DQ}(\lambda_i) \right\rangle \\ \Delta\lambda_i = \varepsilon_m \mathrm{PF}_m, \Delta\mathbf{Y}_m^{DQ}(\lambda_i) = \varepsilon_m \mathbf{Y}_m^{DQ}(\lambda_i) \end{cases} \qquad (7)$$

where, $\mathbf{Z}_{wm}$ represents the submatrix of $\mathbf{Z}_w$ corresponding to rows $2m$-1 to $2m$ and columns $2m$-1 to $2m$. $\operatorname{Res}_{\lambda_i} \mathbf{Z}_{wm}$ denotes the residue of $\mathbf{Z}_{wm}$ in the dynamic system matrix at the eigenvalue $\lambda_i$, while * indicates the conjugate transpose of the matrix. $\langle \cdot \rangle$ represents the Frobenius inner product, $\Delta\lambda_i$ and $\Delta\mathbf{Y}_m^{DQ}$ represent perturbations in $\lambda_i$ and $\mathbf{Y}_m^{DQ}$, respectively. $\varepsilon_m$ is a scalar that represents the degree of perturbation of $\mathbf{Y}_m^{DQ}$.

$\mathbf{Z}_{wm}$ is the equivalent impedance of the whole-system at bus $m$. The equivalent admittance of the rest of the system at the power source $m$ port is denoted as $\mathbf{Y}_{gm}$. According to the circuit relationship in Fig. 1, $(\mathbf{Z}_{wm})^{-1} = \mathbf{Y}_m^{DQ} + \mathbf{Y}_{gm}$ is established. Both $\mathbf{Z}_{wm}$ and $\mathbf{Y}_m^{DQ}$ can be directly obtained through measurement and curve fitting at the $m$-th power source port. The calculation of MAI relies solely on local system information, making it highly scalable. The overall effect of the interactions among various control loops within a power source is captured in the inverter's admittance model. At the same time, the interactions of the power sources and the grid are reflected in $\mathrm{PF}_m$ in (7). Therefore, MAI is an element-level modal analysis method. Unlike the MASS method, MAI cannot identify the dominant control loops within a GFL that influence system stability.

## III. ADMITTANCE DECOMPOSITION IN SYNCHRONOUS DYNAMICS

The control structure of the GFL is shown in Fig. 2. In the following, the subscript $m$ refers to the $m$-th bus in the system. The admittances of all the power sources must be aligned with the global reference frame. In Fig. 3, $\omega_b$ represents the angular velocity of the system's reference source, whereas $\theta_m$, $\theta_{m0}$, and $\Delta\theta_m$ denote the angle of GFL$m$ relative to the reference, the steady-state angle, and the angular deviation, respectively. When the frequency reference signal of the frame change is constant or the PLL is disconnected (i.e. $\Delta\theta_m = 0$), this implies that the swing and stable frame systems coincide [17]. The equivalent admittance $\mathbf{Y}_{me}^{dq}$ from the ED part has been thoroughly discussed in [22]. In this paper, we focus on the decomposition of the overall admittance $\mathbf{Y}_m^{DQ}$ considering the PLL dynamics, as shown in Fig. 4.

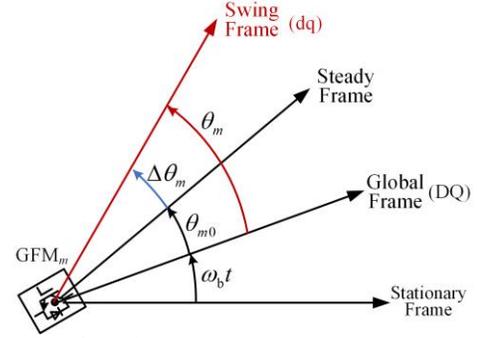

Fig. 3. Frame Transformation.

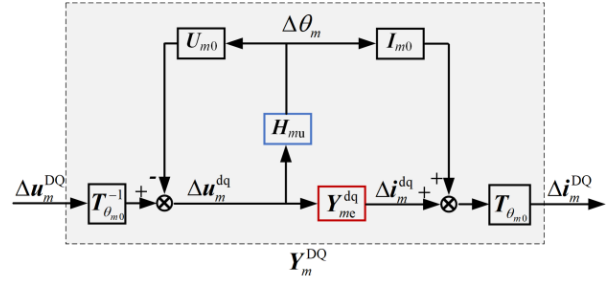

Fig. 4. Admittance Transformation.

The small-signal transfer function of the PLL is given in (8).

$$\begin{cases} \Delta\theta_m = \dfrac{1}{s} \cdot \left( K_{mp}^{PLL} + \dfrac{K_{mi}^{PLL}}{s} \right) \Delta u_m^q \\ \quad = \left( \dfrac{s K_{mp}^{PLL} + K_{mi}^{PLL}}{s^2} \begin{bmatrix} 0 & 1 \end{bmatrix} \right) \begin{bmatrix} \Delta u_m^d \\ \Delta u_m^q \end{bmatrix} \\ \Delta\theta_m = \mathbf{H}_{mu} \Delta\mathbf{u}_m^{dq} \\ \mathbf{H}_{mu} = \dfrac{s K_{mp}^{PLL} + K_{mi}^{PLL}}{s^2} \begin{bmatrix} 0 & 1 \end{bmatrix} \end{cases} \qquad (8)$$

where $K_{mp}^{PLL}$ and $K_{mi}^{PLL}$ represent the proportional and integral gain of the PLL. $\mathbf{H}_{mu}$ is the voltage transfer function.

As shown in Fig. 3, the voltage variables of GFL$m$ can be aligned from the swing frame to the global frame through coordination transformation and then linearized at the equilibrium:

$$\begin{cases} \begin{bmatrix} \Delta u_m^D \\ \Delta u_m^Q \end{bmatrix} = \begin{bmatrix} \cos\theta_{m0} & -\sin\theta_{m0} \\ \sin\theta_{m0} & \cos\theta_{m0} \end{bmatrix} \left( \begin{bmatrix} \Delta u_m^d \\ \Delta u_m^q \end{bmatrix} + \begin{bmatrix} -u_{m0}^q \\ u_{m0}^d \end{bmatrix} \Delta\theta_m \right) \\ \Delta\mathbf{u}_m^{DQ} = \mathbf{T}_{\theta_{m0}} (\Delta\mathbf{u}_m^{dq} + \mathbf{U}_{m0}\Delta\theta_m) \\ \mathbf{T}_{\theta_{m0}} = \begin{bmatrix} \cos\theta_{m0} & -\sin\theta_{m0} \\ \sin\theta_{m0} & \cos\theta_{m0} \end{bmatrix}, \mathbf{U}_{m0} = \begin{bmatrix} -u_{m0}^q \\ u_{m0}^d \end{bmatrix} \end{cases} \qquad (9)$$

where $\mathbf{U}_{m0}$, $u_{m0}^q$, $u_{m0}^d$ denote the voltage at the point of common coupling (PCC), q-axis, and d-axis of the equilibrium, respectively. $\mathbf{T}_{\theta_{m0}}$ is the coordination transformation matrix.

The current follows a similar relationship as in (9). From this, we can derive the overall admittance $\mathbf{Y}_m^{DQ}$ considering the PLL in the global frame, as shown in (10).

$$\mathbf{Y}_m^{DQ} = \mathbf{T}_{\theta_{m0}} (\mathbf{Y}_{me}^{dq} + \mathbf{I}_{m0}\mathbf{H}_{mu}) \cdot (\mathbf{E}_2 + \mathbf{U}_{m0}\mathbf{H}_{mu})^{-1} \mathbf{T}_{\theta_{m0}}^{-1} \qquad (10)$$

where $\boldsymbol{E}_2$ is a second-order identity matrix. $\boldsymbol{I}_{m0}$ denotes the output current of GFL. $\boldsymbol{Y}_{me}^{dq}$ represents the equivalent admittance of ED, as shown in Fig. 4.

$\boldsymbol{U}_{m0}$ and $\boldsymbol{H}_{mu}$ are matrices of order 2×1 and 1×2, respectively, so $(\boldsymbol{E}_2+\boldsymbol{U}_{m0}\boldsymbol{H}_{mu})^{-1}$ can be written as $(\boldsymbol{E}_2+\boldsymbol{U}_{m0}\cdot 1\cdot\boldsymbol{H}_{mu})^{-1}$, which can be simplified using the matrix inversion lemma, as shown in (11).

$$\begin{cases}(\boldsymbol{a}+\boldsymbol{b}\boldsymbol{d}^{-1}\boldsymbol{c})^{-1}=\boldsymbol{a}^{-1}-\boldsymbol{a}^{-1}\boldsymbol{b}(\boldsymbol{d}+\boldsymbol{c}\boldsymbol{a}^{-1}\boldsymbol{b})^{-1}\boldsymbol{c}\boldsymbol{a}^{-1}\\(\boldsymbol{E}_2+\boldsymbol{U}_{m0}\cdot 1\cdot\boldsymbol{H}_{mu})^{-1}=\boldsymbol{E}_2-\boldsymbol{U}_{m0}(1+\boldsymbol{H}_{mu}\boldsymbol{U}_{m0})^{-1}\boldsymbol{H}_{mu}\end{cases}\tag{11}$$

where $\boldsymbol{a},\boldsymbol{b},\boldsymbol{c},$ and $\boldsymbol{d}$ are arbitrary matrices that are dimensionally compatible and satisfy the rules of matrix multiplication. We substitute (11) into (10) to decompose $\boldsymbol{Y}_m^{DQ}$ into four parts, as shown in (12).

$$\boldsymbol{T}_{\theta_{m0}}^{-1}\boldsymbol{Y}_m^{DQ}\boldsymbol{T}_{\theta_{m0}}=\boldsymbol{Y}_{me}^{dq}+\boldsymbol{I}_{m0}\boldsymbol{H}_{mu}-\boldsymbol{Y}_{me}^{dq}\boldsymbol{U}_{m0}(1+\boldsymbol{H}_{mu}\boldsymbol{U}_{m0})^{-1}\boldsymbol{H}_{mu}$$
$$-\boldsymbol{I}_{m0}\boldsymbol{H}_{mu}\boldsymbol{U}_{m0}(1+\boldsymbol{H}_{mu}\boldsymbol{U}_{m0})^{-1}\boldsymbol{H}_{mu}\tag{12}$$

Note that $\boldsymbol{H}_{mu}\boldsymbol{U}_{m0}$ is a scalar. To simplify the expression, we can rewrite $\boldsymbol{I}_{m0}\boldsymbol{H}_{mu}$ as $(1+\boldsymbol{H}_{mu}\boldsymbol{U}_{m0})(1+\boldsymbol{H}_{mu}\boldsymbol{U}_{m0})^{-1}\boldsymbol{I}_{m0}\boldsymbol{H}_{mu}$.

$$\boldsymbol{T}_{\theta_{m0}}^{-1}\boldsymbol{Y}_m^{DQ}\boldsymbol{T}_{\theta_{m0}}=\boldsymbol{Y}_{me}^{dq}+(1+\boldsymbol{H}_{mu}\boldsymbol{U}_{m0})(1+\boldsymbol{H}_{mu}\boldsymbol{U}_{m0})^{-1}\boldsymbol{I}_{m0}\boldsymbol{H}_{mu}$$
$$-(1+\boldsymbol{H}_{mu}\boldsymbol{U}_{m0})^{-1}\boldsymbol{Y}_{me}^{dq}\boldsymbol{U}_{m0}\boldsymbol{H}_{mu}$$
$$-\boldsymbol{H}_{mu}\boldsymbol{U}_{m0}(1+\boldsymbol{H}_{mu}\boldsymbol{U}_{m0})^{-1}\boldsymbol{I}_{m0}\boldsymbol{H}_{mu}$$
$$=\boldsymbol{Y}_{me}^{dq}+(1+\boldsymbol{H}_{mu}\boldsymbol{U}_{m0})^{-1}\boldsymbol{I}_{m0}\boldsymbol{H}_{mu}$$
$$-(1+\boldsymbol{H}_{mu}\boldsymbol{U}_{m0})^{-1}\boldsymbol{Y}_{me}^{dq}\boldsymbol{U}_{m0}\boldsymbol{H}_{mu}\tag{13}$$

Finally (10) is divided into three parts.

$$\boldsymbol{Y}_m^{DQ}=\boldsymbol{T}_{\theta_{m0}}\left(\boldsymbol{Y}_{me}^{dq}+\boldsymbol{Y}_{me}^{dq}\boldsymbol{K}_{mU}/U_m^{PLL}+\boldsymbol{K}_{mI}/U_m^{PLL}\right)\boldsymbol{T}_{\theta_{m0}}^{-1}$$
$$where\begin{cases}\boldsymbol{K}_{mU}=\begin{bmatrix}0 & u_{m0}^q\\0 & -u_{m0}^d\end{bmatrix}\boldsymbol{K}_{mI}=\begin{bmatrix}0 & -i_{m0}^q\\0 & i_{m0}^d\end{bmatrix}\\U_m^{PLL}=u_{m0}^d+s^2/(K_{mp}^{PLL}+sK_{mp}^{PLL})\\u_{m0}^d=U_m,u_{m0}^q=0\\i_{m0}^d=P_m/U_m,i_{m0}^q=-Q_m/U_m\end{cases}\tag{14}$$

where $\boldsymbol{K}_{mU}$ and $\boldsymbol{K}_{mI}$ are coefficient matrices related to steady-state voltage $u_{m0}^q$, $u_{m0}^d$ and current $i_{m0}^q$, $i_{m0}^d$, respectively. $U_m^{PLL}$ represents the equivalent voltage associated with the PLL parameters. $U_m$, $P_m$, and $Q_m$ are the voltage, active and reactive power measured at the PCC, respectively.

The first and third parts of $\boldsymbol{Y}_m^{DQ}$ are solely related to the CCL and PLL, respectively, while the second part is influenced by the coupling effect between the two control loops. Therefore, the decomposition of $\boldsymbol{Y}_m^{DQ}$ can be rewritten as (15).

$$\boldsymbol{Y}_m^{DQ}=\boldsymbol{Y}_{me}^{DQ}+\boldsymbol{Y}_{me,s}^{DQ}+\boldsymbol{Y}_{ms}^{DQ}\tag{15}$$

## IV. Extended Impedance Modal Analysis

### A. Overall Participation Factors of Different Dynamics

Unlike the overall participation factor $PF_m$ of GFL$m$ in (7), this section derives the respective overall participation factors of the internal electromagnetic and synchronous dynamics of GFL$m$. Since the MAI method has been Extended from the overall dynamic participation assessment at the element level to internal dynamic participation assessment within the GFL, the proposed method is referred to as extended modal analysis based on the impedance model (EMAI).

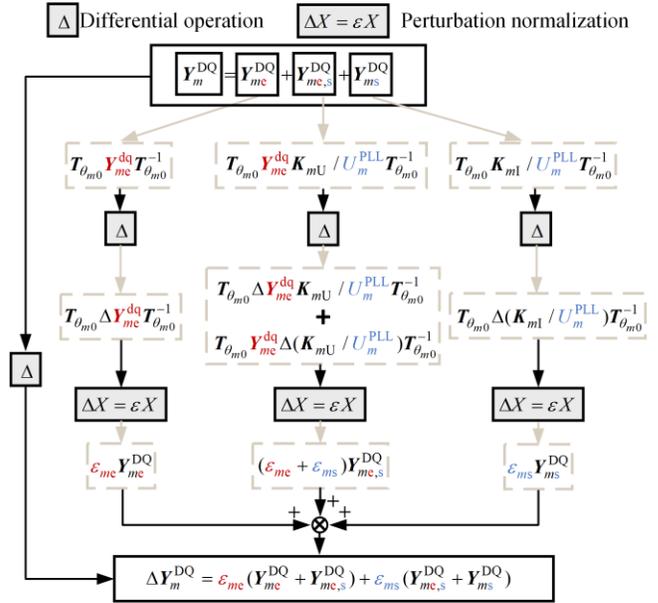

Fig. 5 Schematic diagram of the admittance differential operation.

Taking the perturbation of (15), the schematic diagram of the computation process is shown in Fig. 5.

$$\begin{cases}\Delta\boldsymbol{Y}_m^{DQ}=\varepsilon_{me}(\boldsymbol{Y}_{me}^{DQ}+\boldsymbol{Y}_{me,s}^{DQ})+\varepsilon_{ms}(\boldsymbol{Y}_{ms}^{DQ}+\boldsymbol{Y}_{me,s}^{DQ})\\\Delta\boldsymbol{Y}_{me}^{dq}=\varepsilon_{me}\boldsymbol{Y}_{me}^{dq},\Delta(1/U_m^{PLL})=\varepsilon_{ms}(1/U_m^{PLL})\end{cases}\tag{16}$$

where $\varepsilon_{me}$ and $\varepsilon_{ms}$ represent the degree of perturbation in $\Delta\boldsymbol{Y}_{me}^{dq}$ and $\Delta(1/U_m^{PLL})$, respectively.

Combining (7) and (16) offers deeper insight into identifying the primary control loops that most significantly impact the system. Analogous to (7), we define $PF_{me1}$ and $PF_{ms1}$ in (17) to characterize the participation of ED and SD, respectively. The subscript 1 indicates results computed using EMAI.

$$\begin{cases}\Delta\lambda_i=\left\langle-\mathrm{Res}_{\lambda_i}^*\boldsymbol{Z}_{wm},\Delta\boldsymbol{Y}_m^{DQ}(\lambda_i)\right\rangle=\varepsilon_{me}PF_{me1}+\varepsilon_{ms}PF_{ms1}\\PF_{me1}=\left\langle-\mathrm{Res}_{\lambda_i}^*\boldsymbol{Z}_{wm},\boldsymbol{Y}_{me}^{DQ}(\lambda_i)+\boldsymbol{Y}_{me,s}^{DQ}(\lambda_i)\right\rangle\\PF_{ms1}=\left\langle-\mathrm{Res}_{\lambda_i}^*\boldsymbol{Z}_{wm},\boldsymbol{Y}_{ms}^{DQ}(\lambda_i)+\boldsymbol{Y}_{ms}^{DQ}(\lambda_i)\right\rangle\end{cases}\tag{17}$$

where $PF_{me1}$ and $PF_{ms1}$ reflect the equivalent overall PF of the ED and SD, respectively. Notably, $PF_{me1}$ is jointly influenced by $\boldsymbol{Y}_{me}^{DQ}$ and $\boldsymbol{Y}_{me,s}^{DQ}$, whereas $PF_{ms1}$ results from the interaction of $\boldsymbol{Y}_{me,s}^{DQ}$ and $\boldsymbol{Y}_{ms}^{DQ}$.

### B. Overall Participation Evaluation of Different Dynamics

The participation of ED and SD can also be evaluated from the state variable PF using the MASS method. The subscript 2 indicates results computed using MASS.

$$\begin{cases}PF_{me2}=PF_{mi}^d+PF_{mi}^q+PF_{miPI}^d+PF_{miPI}^d\\PF_{ms2}=PF_{m\theta}+PF_{m\omega}\end{cases}\tag{18}$$

where $PF_{mi}^d$, $PF_{mi}^q$, $PF_{miPI}^d$, $PF_{miPI}^d$, $PF_{m\theta}$, and $PF_{m\omega}$ represent the PF of the dq-axis inductive current, the PF of the dq-axis PI controller integrator in the CCL, the PF of the PLL's angle variable, and the PF of the PI integrator in the PLL, respectively. They can be obtained by (4).

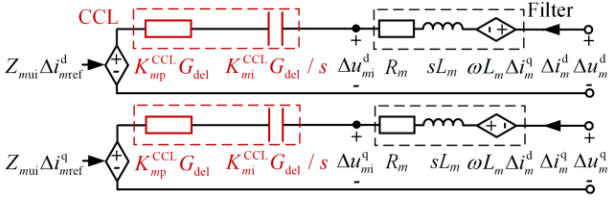

Fig. 6 The equivalent circuit of CCL.

To evaluate the ED and SD participation degree of different GFLs, the participation ratio (PR) is defined to normalize the PF from the MAI, EMAI, and MASS methods, as shown in (19). The subscript $l$=1 or 2, indicates PF or PR computed from the EMAI and MASS method, respectively.

$$
\begin{cases}
\mathrm{PR}_m = \mid \mathrm{PF}_m^{\mathrm{PLL}} \mid / \sum_{m=1}^{n} \mid \mathrm{PF}_m \mid \\
\mathrm{PR}_{mel} = \mid \mathrm{PF}_{mel} \mid / \sum_{m=1}^{n} (\mid \mathrm{PF}_{mel} \mid + \mid \mathrm{PF}_{msl} \mid) \\
\mathrm{PR}_{msl} = \mid \mathrm{PF}_{msl} \mid / \sum_{m=1}^{n} (\mid \mathrm{PF}_{mel} \mid + \mid \mathrm{PF}_{msl} \mid)
\end{cases}
\tag{19}
$$

where $\mid \ \mid$ indicates the absolute value.

### C. Explicit Parameter Participation Factor

Additionally, $\partial \boldsymbol{Y}_m^{\mathrm{DQ}} / \partial \rho$, the sensitivity of the equivalent admittance to any specific parameter can be used to identify the dominant factors affecting system dynamics. The PPF, defined in (20) in previous work [18]-[19], is difficult to obtain analytically. Previous research used numerical differences as a substitute for computing the partial derivatives, which requires a small step size $\Delta \rho$ and may result in unexpected errors if not carefully managed.

$$
\begin{cases}
\Delta \lambda_i = \left\langle -\operatorname{Res}_{\lambda_i}^* \boldsymbol{Z}_{wm}, \dfrac{\partial \boldsymbol{Y}_m^{\mathrm{DQ}}(\lambda_i)}{\partial \rho} \right\rangle \Delta \rho = \mathrm{PPF}_{m,\rho} \Delta \rho \\
\dfrac{\partial \boldsymbol{Y}_m^{\mathrm{DQ}}(\lambda_i)}{\partial \rho} \approx \dfrac{\Delta \boldsymbol{Y}_{m,\Delta\rho}^{\mathrm{DQ}}(\lambda_i) - \Delta \boldsymbol{Y}_m^{\mathrm{DQ}}(\lambda_i)}{\Delta \rho}
\end{cases}
\tag{20}
$$

Based on the admittance decomposition method proposed in Section III, we present an explicit expression for $\partial \boldsymbol{Y}_m^{\mathrm{DQ}} / \partial \rho$. We begin by considering the parameters of CCL. As shown in Fig. 5, $\Delta \boldsymbol{Y}_m^{\mathrm{DQ}}$ can be expressed as:

$$
\Delta \boldsymbol{Y}_m^{\mathrm{DQ}} = \boldsymbol{T}_{\theta_{n0}} \Delta \boldsymbol{Y}_m^{\mathrm{dq}} (\boldsymbol{E}_2 + \boldsymbol{K}_{m\mathrm{U}} / U_m^{\mathrm{PLL}}) \boldsymbol{T}_{\theta_{n0}}^{-1}
\tag{21}
$$

All controls in CCL can be converted to impedance, either in series or parallel with the filter circuit. For instance, when voltage feedforward, virtual impedance, and cross decoupling are all set to zero, the PI controller of CCL can be used as a reference point to understand this conversion process.

As shown in Fig. 6, $\Delta u_{mi}^{\mathrm{d}}$ and $\Delta u_{mi}^{\mathrm{q}}$ are the d-axis and q-axis terminal voltages of GFL $m$. $\Delta i_{mref}^{\mathrm{d}}$ and $\Delta i_{mref}^{\mathrm{q}}$ are the d-axis and q-axis current reference value of CCL, and they are both 0 if there is no external control loop affecting them (such as the DC voltage loop). $Z_{mui}=(K_{mp}^{\mathrm{CCL}} + K_{mpi}^{\mathrm{CCL}} / s) e^{-1.5 T_s s}$ is the transfer function of the current to the controlled voltage. $T_s$ represents the sampling period. $K_{mp}^{\mathrm{CCL}}$ and $K_{mi}^{\mathrm{CCL}}$ are CCL proportional gain and integral gain.

Since the PI controller of the CCL can be represented as an impedance in series with the filter inductor [22], the admittance perturbation can be converted into an impedance perturbation.

$$
\Delta \boldsymbol{Y}_{me}^{\mathrm{dq}} = \frac{\partial \boldsymbol{Y}_{me}^{\mathrm{dq}}}{\partial \rho} \Delta \rho = -\boldsymbol{Y}_{me}^{\mathrm{dq}} \frac{\partial \boldsymbol{Z}_{me}^{\mathrm{dq}}}{\partial \rho} \boldsymbol{Y}_{me}^{\mathrm{dq}} \Delta \rho
\tag{22}
$$

The impedance decomposition of the CCL in [22] directly yields $\partial \boldsymbol{Z}_{me}^{\mathrm{dq}} / \partial \rho$. To be specific, the parameters related to the CCL include the proportional gain $K_{mp}^{\mathrm{CCL}}$ and the integral gains $K_{mi}^{\mathrm{CCL}}$ of the PI controller, as shown in (23).

$$
\frac{\partial \boldsymbol{Z}_{me}^{\mathrm{dq}}}{\partial K_{mp}^{\mathrm{CCL}}} = \begin{bmatrix} 1 & 0 \\ 0 & 1 \end{bmatrix} e^{-1.5 T_s s}, \frac{\partial \boldsymbol{Z}_{me}^{\mathrm{dq}}}{\partial K_{mi}^{\mathrm{CCL}}} = \begin{bmatrix} 1/s & 0 \\ 0 & 1/s \end{bmatrix} e^{-1.5 T_s s}
\tag{23}
$$

where $G_{del}$ represents the delay function.

For the PLL parameters, the sensitivity is solely related to the scalar $U_m^{\mathrm{PLL}}$ in (24).

$$
\Delta \boldsymbol{Y}_m^{\mathrm{DQ}} = \Delta (1/U_m^{\mathrm{PLL}}) \boldsymbol{T}_{\theta_{n0}} (\boldsymbol{Y}_{me}^{\mathrm{dq}} \boldsymbol{K}_{m\mathrm{U}} + \boldsymbol{K}_{mI}) \boldsymbol{T}_{\theta_{n0}}^{-1}
\tag{24}
$$

The sensitivity of $K_{mp}^{\mathrm{PLL}}$ and $K_{mi}^{\mathrm{PLL}}$ is shown in (25).

$$
\begin{cases}
\dfrac{\partial \boldsymbol{Y}_m^{\mathrm{DQ}}}{\partial K_{mp}^{\mathrm{PLL}}} = \dfrac{s \boldsymbol{T}_{\theta_{n0}} (\boldsymbol{Y}_{me}^{\mathrm{dq}} \boldsymbol{K}_{mU} + \boldsymbol{K}_{mI}) \boldsymbol{T}_{\theta_{n0}}^{-1}}{(U_m^{\mathrm{PLL}})^2 (K_{mi}^{\mathrm{PLL}} + s K_{mp}^{\mathrm{PLL}})^2} \\
\dfrac{\partial \boldsymbol{Y}_m^{\mathrm{DQ}}}{\partial K_{mi}^{\mathrm{PLL}}} = \dfrac{\boldsymbol{T}_{\theta_{n0}} (\boldsymbol{Y}_{me}^{\mathrm{dq}} \boldsymbol{K}_{mU} + \boldsymbol{K}_{mI}) \boldsymbol{T}_{\theta_{n0}}^{-1}}{(U_m^{\mathrm{PLL}})^2 (K_{mi}^{\mathrm{PLL}} + s K_{mp}^{\mathrm{PLL}})^2}
\end{cases}
\tag{25}
$$

The PPF of PLL and CCL can be obtained by combining (20)-(25).

### D. Influence of External Control Loops

When the DC voltage loop (DVL) is considered (i.e., $\Delta i_{mref}^{\mathrm{d}} \neq 0$), the effect of DVL can be equivalent to an additional branch of the d-axis. At this time, the dynamic of the DC voltage loop is also included in ED, so the admittance $\boldsymbol{Y}_{me}^{\mathrm{dq}}$ shown in Fig. 6 needs to be corrected.

The power equation of the DC side is as follows.

$$
P_m^{\mathrm{DC}} + C_m^{\mathrm{DC}} U_m^{\mathrm{DC}} \dot{U}_m^{\mathrm{DC}} = u_m^{\mathrm{d}} i_m^{\mathrm{d}} + u_m^{\mathrm{q}} i_m^{\mathrm{q}}
\tag{26}
$$

where $P_m^{\mathrm{DC}}$, $C_m^{\mathrm{DC}}$, and $U_m^{\mathrm{DC}}$ are the DC active power, capacitance, and voltage. $\dot{U}_m^{\mathrm{DC}}$ is the differential of $U_m^{\mathrm{DC}}$.

Linearize (26) and note that $\Delta P_m^{\mathrm{DC}} = 0$ and $\Delta U_m^{\mathrm{DC}} = 0$ hold.

$$
C_m^{\mathrm{DC}} s U_{m0}^{\mathrm{DC}} \Delta U_m^{\mathrm{DC}} = u_{mi0}^{\mathrm{d}} \Delta i_m^{\mathrm{d}} + u_{m0}^{\mathrm{q}} \Delta i_m^{\mathrm{q}} + i_{m0}^{\mathrm{d}} \Delta u_m^{\mathrm{d}} + i_{m0}^{\mathrm{q}} \Delta u_{mi}^{\mathrm{q}}
\tag{27}
$$

The equation of DC voltage is as follows.

$$
i_{mref}^{\mathrm{d}} = (K_{mp}^{\mathrm{DVL}} + K_{mi}^{\mathrm{DVL}} / s)(U_m^{\mathrm{DC}} - U_{mref}^{\mathrm{DC}})
\tag{28}
$$

where $K_{mp}^{\mathrm{DVL}}$ and $K_{mi}^{\mathrm{DVL}}$ are DVL proportional gain and integral gain. $U_{mref}^{\mathrm{DC}} = U_{m0}^{\mathrm{DC}}$ is the reference value of $U_m^{\mathrm{DC}}$.

Linearize (28).

$$
\Delta i_{mref}^{\mathrm{d}} = (K_{mp}^{\mathrm{DVL}} + \frac{K_{mi}^{\mathrm{DVL}}}{s}) \Delta U_m^{\mathrm{DC}}
\tag{29}
$$

Combination (27) and (29).

$$
\Delta i_{mref}^{\mathrm{d}} = \frac{u_{mi0}^{\mathrm{d}} \Delta i_m^{\mathrm{d}} + u_{m0}^{\mathrm{q}} \Delta i_m^{\mathrm{q}} + i_{m0}^{\mathrm{d}} \Delta u_m^{\mathrm{d}} + i_{m0}^{\mathrm{q}} \Delta u_{mi}^{\mathrm{q}}}{C_m^{\mathrm{DC}} s U_{m0}^{\mathrm{DC}}}.
$$
$$
(K_{mp}^{\mathrm{DVL}} + K_{mi}^{\mathrm{DVL}} / s)
$$
$$
= K_{mi}^{\mathrm{dd}} \Delta i_m^{\mathrm{d}} + K_{mi}^{\mathrm{dq}} \Delta i_m^{\mathrm{q}} + K_{miu}^{\mathrm{dd}} \Delta u_m^{\mathrm{d}} + K_{miu}^{\mathrm{dq}} \Delta u_{mi}^{\mathrm{q}}
\tag{30}
$$

where $K_{mi}^{\mathrm{dd}}$, $K_{mi}^{\mathrm{dq}}$, $K_{miu}^{\mathrm{dd}}$, and $K_{miu}^{\mathrm{dq}}$ are the coefficients related to the voltage loop and the steady-state operating point.

According to (30), the equivalent circuit of DVL in d axis can be obtained, as shown in Fig. 7. $Z_s$ and $Y_p$ are DVL equivalent series impedance and parallel admittance, respectively.

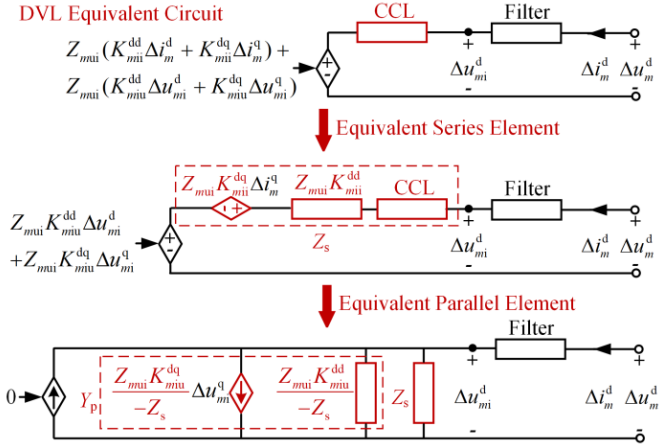

Fig. 7 The equivalent circuit of DVL.

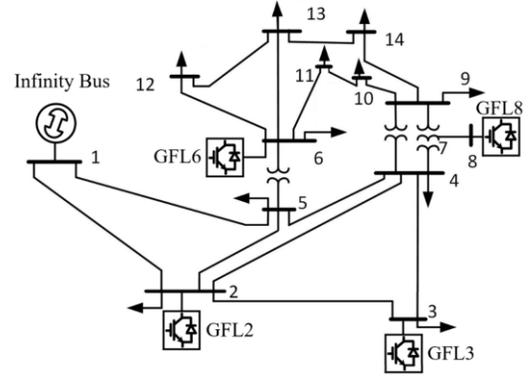

Fig. 8 Modified 14 bus System topology.

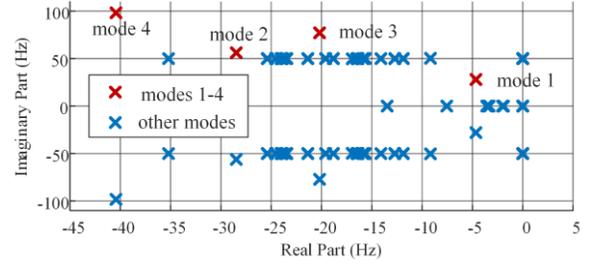

Fig. 9 Modified 14 bus System Pole diagram.

TABLE I
COMPARISON OF DIFFERENT METHODS

| Methods | MASS | MAI | EMAI |
|---|---|---|---|
| Applicability to systems with black-box or grey-box models | - | + | + |
| Calculations can require only local system's information | - | + | + |
| Determine the dominant element | + | + | + |
| Determine the key dynamics of the dominant element | + | - | + |

Similarly, PPF for calculating proportional gain $K_{mp}^{\mathrm{DVL}}$ and integral gain $K_{mi}^{\mathrm{DVL}}$ in DVL can also be obtained according to (30) and Fig. 7.

Therefore, considering the influence of DVL, $\mathrm{PF}_{me2}$ of (18) should be corrected to (31).

$$\begin{cases} \mathrm{PF}_{me2} = \mathrm{PF}_{mi}^{d} + \mathrm{PF}_{mi}^{q} + \mathrm{PF}_{miPI}^{d} + \mathrm{PF}_{miPI}^{q} + \mathrm{PF}_{mu}^{DC} + \mathrm{PF}_{muPI}^{DC} \\ \mathrm{PF}_{ms2} = \mathrm{PF}_{m\theta} + \mathrm{PF}_{m\omega} \end{cases} \quad (31)$$

where $\mathrm{PF}_{mu}^{DC}$ and $\mathrm{PF}_{muPI}^{DC}$ represent the PF of the DC voltage and integrator of PI controller in DVL.

### E. Comparison of Different Methods

This paper decomposes the admittance of GFL to identify the participation factors representing its internal dynamics and proposes the EMAI method. EMAI effectively addresses the limitations of MAI in evaluating the internal dynamics of GFL. The comparative results of the three different modal analysis methods, MASS, MAI, and EMAI, are concluded in Table I.

## V. CASE STUDIES

This section evaluates the performance of the proposed EMAI method using two systems of different scales. A modified IEEE 14-bus system is used to validate EMAI's ability to accurately capture the dominant dynamics within GFLs under various conditions. A modified IEEE 68-bus system is employed to demonstrate EMAI's applicability in large-scale power systems and to evaluate the effectiveness of the proposed PPF in guiding improvements to system damping. All simulation models and data were generated using the open-source software package *Simplus Grid Tool* [23].

### A. 14 bus System

The modified 14-bus system is shown in Fig. 8, where Bus 1 is configured as the infinite bus, and Buses 2, 3, 6, and 8 are connected to GFLs. The parameters for all GFLs are listed in Table II (without DVL), and the system pole distribution is depicted in Fig. 9. The base frequency of the 14-bus system is 50 Hz. Modes 1, 2, 3, and 4 correspond to different frequency modes, with the following values: $2\pi(-4.65+j27.98)$, $2\pi(-28.47+j56.16)$, $2\pi(-20.19+j77.11)$, and $2\pi(-40.43+j98.24)$, respectively.

The PR calculation results from different methods are shown in Fig. 10. For example, in Mode 1, GFL8 exhibits the highest ED participation among all GFLs, so it is referred to as "GFL 8 dominant ED" in the following text and figures. The nearly identical bar heights for different modes in Fig. 10 indicate that all three methods provide consistent assessments of the overall participation of various GFLs. However, MAI fails to capture the dominant dynamics within individual GFLs. In contrast, the proposed EMAI method not only identifies the interactions between each GFL and the grid but also quantifies the complex coupling interactions among different dynamics within each GFL. The results from the EMAI and MASS methods are highly similar, further validating the effectiveness of the EMAI method.

TABLE II
14 BUS SYSTEM PARAMETERS

| Parameter | Symbol | GFL6 | Other GFL |
|---|---|---|---|
| CCL Proportional Gain | $K_{p}^{CCL}$ | 0.24 | 0.24 |
| CCL Integral Gain | $K_{i}^{CCL}$ | 150.79 | 150.79 |
| PLL Proportional Gain | $K_{p}^{PLL}$ | 125.66 | 31.42 |
| PLL Integral Gain | $K_{i}^{PLL}$ | 3947.84 | 246.74 |
| AC L Filter | $R$(p.u.) | 0.01 | 0.01 |
| AC L Filter | $L$(p.u.) | 0.03 | 0.03 |

TABLE III
THE RELEVANT PARAMETERS OF DVL

| Parameter | Symbol | All GFL |
|---|---|---|
| DVL Proportional Gain | $K_{mp}^{DVL}$ | 98.17 |
| DVL Integral Gain | $K_{mi}^{DVL}$ | 771.06 |
| DC capacitance | $C_{m}^{DC}$ (p.u.) | 1.25 |
| DC voltage reference value | $U_{mref}^{DC}$ (p.u.) | 2.5 |

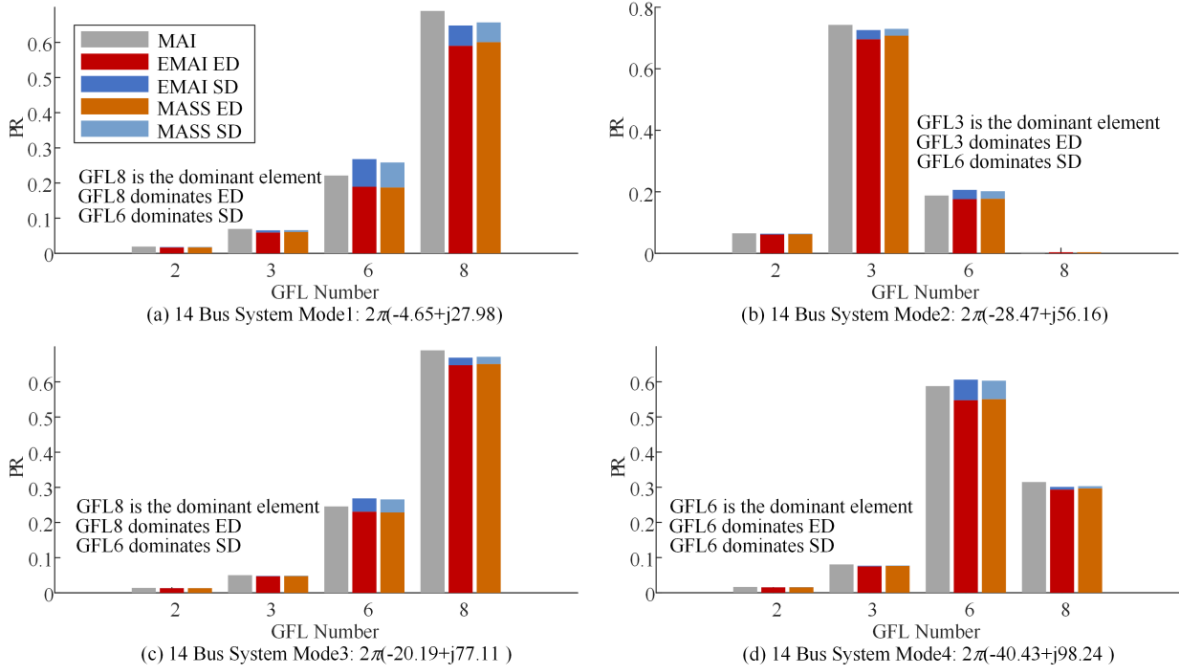

Fig. 10 PR diagrams of 14 bus system (without DVL). (a) Mode1. (b) Mode2. (c) Mode3. (d) Mode4.

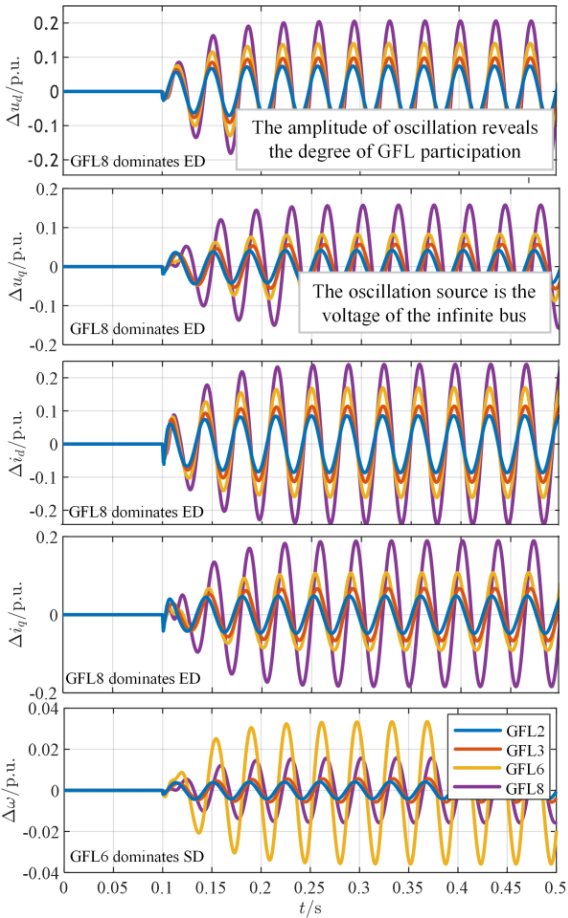

Fig. 11 All GFL waveforms after forced oscillations with infinite bus voltages of 0.05p.u. amplitude and the same frequency as mode 1.

Mode 1 exhibits the weakest damping among the four modes, making it more susceptible to excite oscillations. Fig. 8 shows that GFL8 has the largest electrical distance from the infinite bus compared to the other GFLs. Due to weak grid instability, GFL8 is expected to exhibit higher participation in Mode 1, with disturbance at the infinite bus having the greatest impact on GFL8. Fig. 10(a) indicates that GFL8 has the highest overall participation in Mode 1, primarily due to its high ED proportion. Notably, GFL6 shows the highest SD proportion, which is largely attributed to the large proportional and integral gains of the PLL in GFL6, as shown in Table II.

At 0.1 seconds, a forced oscillation with a magnitude of 0.05 p.u. and a frequency matching Mode 1 is applied to bus 1, resulting in the oscillation waveforms of the GFLs shown in Fig. 11. The largest oscillation amplitudes in the voltage and current waveforms of GFL8 indicate its dominance in ED, while the largest frequency waveform oscillation amplitudes of GFL6 suggest its dominance in SD. The simulation results align with the conclusions in Fig. 10(a), demonstrating that the EMAI method accurately captures the dominant dynamics within different GFL dynamics, thereby confirming its effectiveness.

In addition, when DVL is attached to each GFL of the system, The relevant parameters of DVL are shown in Table III. The four modes in Fig. 10 change when considering DVL. The GFL dynamic participation results are shown in Fig. 12. Obviously, when considering the external control loop, the calculation results of EMAI method are almost the same as that of MASS method, and it has strong robustness.

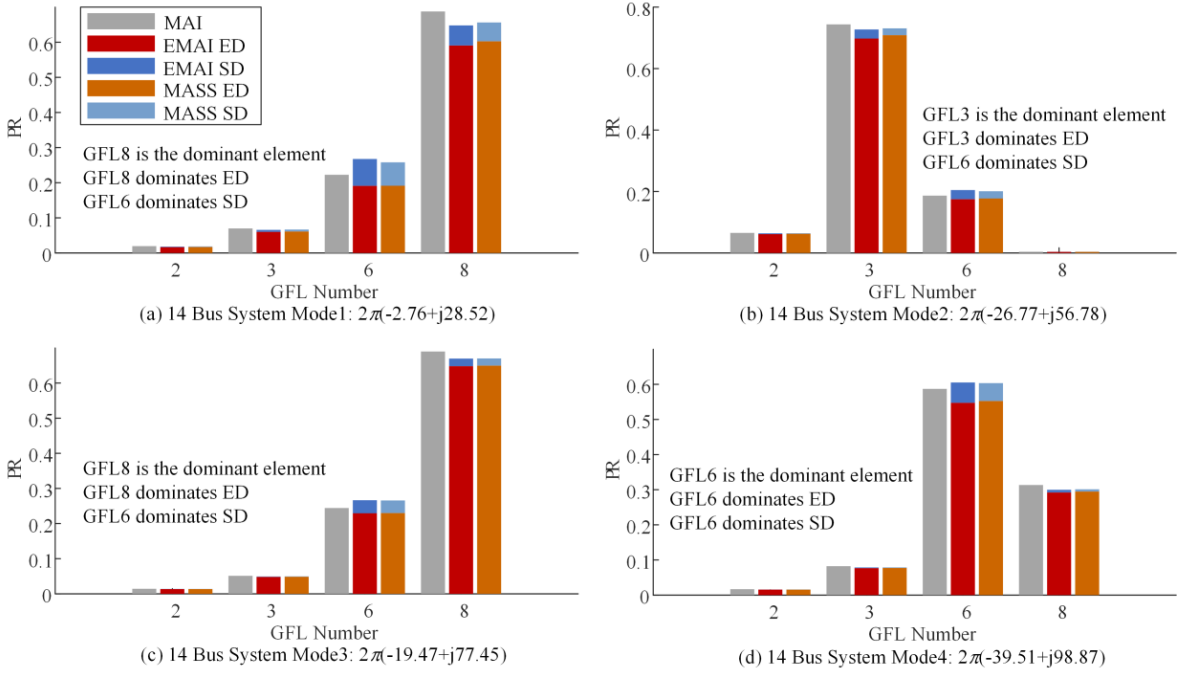

Fig. 12 PR diagrams of 14 bus system (with DVL). (a) Mode1. (b) Mode2. (c) Mode3. (d) Mode4.

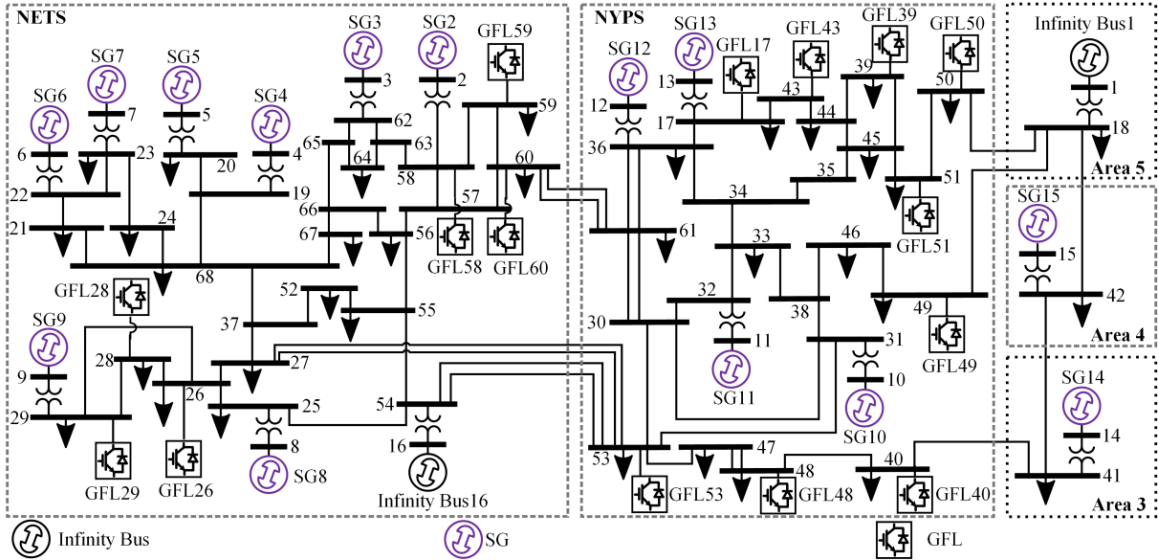

Fig. 13 Modified 68 bus System.

## B. 68 bus System

Fig. 13 presents the topology of the modified 68-bus system, where buses 1 and 16 are designated as infinite buses, buses 2 through 15 are connected to synchronous machines, and the remaining GFLs are distributed across various buses. As shown in Table IV (without DVL), all GFLs share identical parameters. The base frequency of the 68-bus system is 60 Hz. Fig. 14 illustrates the participation evaluation results for four modes of the 68-bus system. The bar charts of PR indicate that the overall participation results of the three methods are nearly identical. Furthermore, the EMAI and MASS methods yield similar evaluations of the participation of various dynamics within GFLs, validating the proposed method's applicability in large-scale power systems.

TABLE IV
68 BUS SYSTEM PARAMETERS

| Parameter | Symbol | All GFL |
|---|---|---|
| CCL Proportional Gain | $K_p^{CCL}$ | 0.06 |
| CCL Integral Gain | $K_i^{CCL}$ | 11.31 |
| PLL Proportional Gain | $K_p^{PLL}$ | 31.42 |
| PLL Integral Gain | $K_i^{PLL}$ | 246.74 |
| AC L Filter | $R$(p.u.) | 0.01 |
| AC L Filter | $L$(p.u.) | 0.03 |

In the participation evaluation results of Mode 1, $2\pi(-5.36 + j10.26)$, shown in Fig. 14(a), GFL40 and GFL48 exhibit high

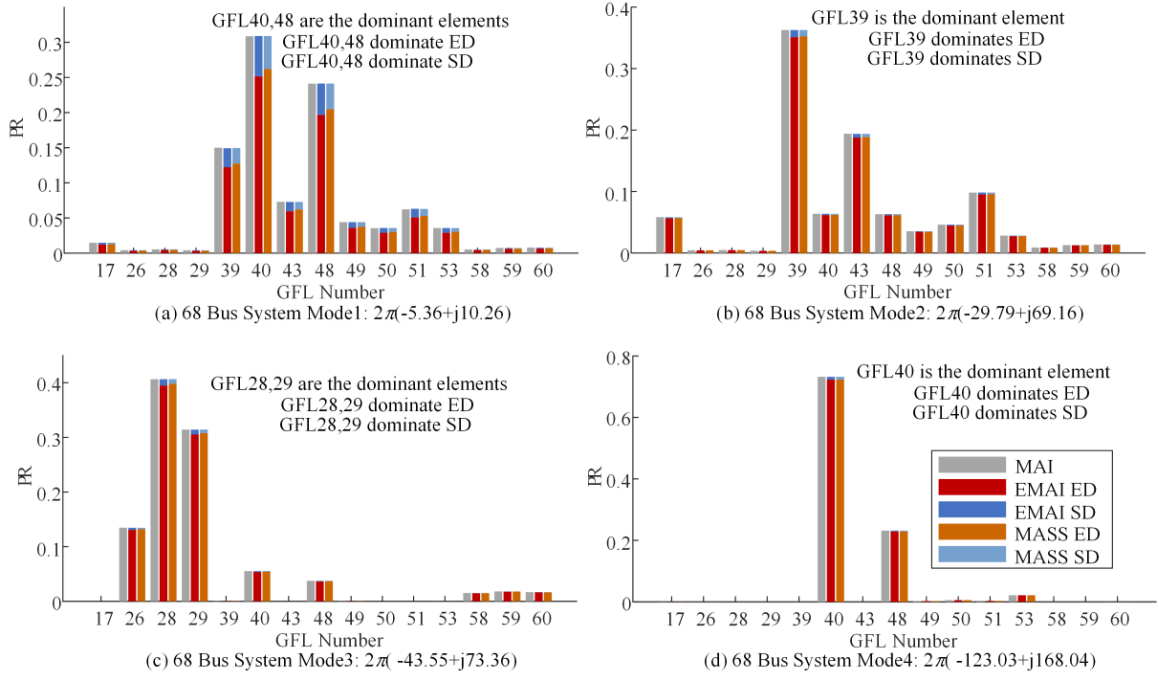

Fig. 14 PR diagrams of 68 bus system. (a) Mode1. (b) Mode2. (c) Mode3. (d) Mode4.

TABLE V
68 BUS SYSTEM PPF

| Symbol | GFL40 | GFL48 |
|---|---|---|
| $K_p^{\text{CCL}}$ | 33.3644-j213.3809 | 4.3156+j168.6776 |
| $K_i^{\text{CCL}}$ | -2.8124+j0.9515 | -2.0828+j1.0209 |
| $K_p^{\text{PLL}}$ | -0.0279+j0.1987 | -0.0002+j0.1563 |
| $K_i^{\text{PLL}}$ | 0.0026-j0.0009 | 0.0019-j0.0009 |

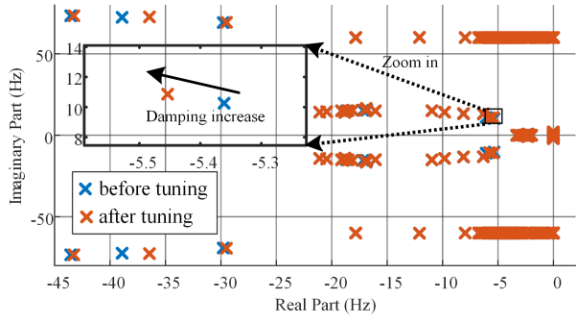

Fig. 15. Pole diagram before and after parameter tuning of 68 bus system.

PR values, suggesting that tuning the control parameters of GFL40 and GFL48 could enhance the damping of this mode. The PPF for GFL40 and GFL48 are listed in Table V. From the real part of PPF, it can be inferred that reducing the proportional gain of the CCL PI controller and increasing the proportional gain of the PLL in GFL40 are effective stability enhancement strategies. Fig. 15 depicts the pole diagram when the proportional gain of the CCL PI controller in GFL40 is reduced from 0.06 to 0 .05. The increased damping of Mode 1 demonstrates the effectiveness of the proposed PPF in guiding system stability improvements.

## VI. CONCLUSION

MASS requires complete system information and is not suitable for black-box or gray-box models. While MAI assesses the overall participation of the GFLs, it lacks deeper insights into the internal dynamics within GFLs. The proposed EMAI method combines the strength of both MASS and MAI, enabling a unified examination of the internal dynamics within GFLs in the system. EMAI provides more detailed insights into system stability, helping to identify the root causes of instability and measure the participation of dynamics of different control loops. Simulation results verify the reliability of the proposed EMAI method. Since EMAI only requires the local admittance model of the system for computation, it exhibits strong scalability, making it suitable for large-scale power systems. Future work will focus on extending this approach to grid-forming inverters.